\documentclass[conference]{IEEEtran}
\IEEEoverridecommandlockouts
\usepackage{cite}
\usepackage{amsmath,amssymb,amsfonts}
\usepackage{algorithmic}
\usepackage{graphicx}
\usepackage{textcomp}
\usepackage{xcolor}
\usepackage{listings}
\usepackage{subfigure}
\usepackage{tikz}
\usepackage{textcomp}
\usepackage{hyperref}
\def\BibTeX{{\rm B\kern-.05em{\sc i\kern-.025em b}\kern-.08em
    T\kern-.1667em\lower.7ex\hbox{E}\kern-.125emX}}

\makeatletter
\newcommand{\linebreakand}{%
  \end{@IEEEauthorhalign}
  \hfill\mbox{}\par
  \mbox{}\hfill\begin{@IEEEauthorhalign}
}

\newcommand\submittedtext{%
  \footnotesize This work has been submitted to the IEEE for possible publication. Copyright may be transferred without notice, after which this version may no longer be accessible.}

\newcommand\submittednotice{%
\begin{tikzpicture}[remember picture,overlay]
\node[anchor=south,yshift=10pt] at (current page.south) {\fbox{\parbox{\dimexpr0.65\textwidth-\fboxsep-\fboxrule\relax}{\submittedtext}}};
\end{tikzpicture}%
}

\makeatother
\begin{document}

\title{Are requirements really all you need? \\ A case study of LLM-driven configuration code generation for automotive simulations}

\author{\IEEEauthorblockN{Krzysztof Lebioda\IEEEauthorrefmark{1}}
\IEEEauthorblockA{email: krzysztof.lebioda@tum.de \\
orcid: 0000-0002-7905-8103}
\and
\IEEEauthorblockN{Nenad Petrovic\IEEEauthorrefmark{1}}
\IEEEauthorblockA{email: nenad.petrovic@tum.de \\
orcid: 0000-0003-2264-7369}
\and
\IEEEauthorblockN{Fengjunjie Pan\IEEEauthorrefmark{1}}
\IEEEauthorblockA{email: f.pan@tum.de \\
orcid: 0009-0005-8303-1156}
\linebreakand
\IEEEauthorblockN{Vahid Zolfaghari\IEEEauthorrefmark{1}}
\IEEEauthorblockA{email: v.zolfaghari@tum.de \\
orcid: 0009-0004-0039-6014}
\and
\IEEEauthorblockN{André Schamschurko\IEEEauthorrefmark{1}}
\IEEEauthorblockA{email: andre.schamschurko@tum.de \\
orcid: 0009-0000-7030-0955}
\and
\IEEEauthorblockN{Alois Knoll\IEEEauthorrefmark{1}}
\IEEEauthorblockA{email: k@tum.de \\
orcid: 0000-0003-4840-076X}
\linebreakand
\IEEEauthorblockA{\IEEEauthorrefmark{1}Technical University of Munich (TUM) \\
School of Computation, Information and Technology (CIT) \\
Chair of Robotics, Artificial Intelligence and Embedded Systems}
}

\maketitle
\submittednotice

\begin{abstract}
Large Language Models (LLMs) are taking many industries by storm.
They possess impressive reasoning capabilities and are capable of handling complex problems, as shown by their steadily improving scores on coding and mathematical benchmarks.
However, are the models currently available truly capable of addressing real-world challenges, such as those found in the automotive industry?
How well can they understand high-level, abstract instructions?
Can they translate these instructions directly into functional code, or do they still need help and supervision?
In this work, we put one of the current state-of-the-art models to the test.
We evaluate its performance in the task of translating abstract requirements, extracted from automotive standards and documents, into configuration code for CARLA simulations.

\end{abstract}


\section{Introduction}

Since the introduction of the groundbreaking ChatGPT-3, Large Language Models (LLMs) show an impressive ability to understand and process natural language, which keeps increasing with each model generation.
It was only natural that these models were soon adopted to help us with many problems involving formal and informal language.
From intent classification, text summarizing, to translating textual descriptions into code, LLMs are having an impact on many industries, such as healthcare \cite{HealthAI, cascellaEvaluatingFeasibilityChatGPT2023} or finances \cite{zhaoRevolutionizingFinanceLLMs2024}, as well as our social lives \cite{radivojevicHumanPerceptionLLMgenerated2024, yangSocialMindLLMbasedProactive2024}.
One of the most affected fields is broadly understood computer science.
Generation of code from a description in natural language using LLMs is currently a hot research topic.
However, most of the work focuses on solving benchmark problems \cite{paulBenchmarksMetricsEvaluations2024}, often not representative of real-world applications.
The literature that evaluates LLMs in handling real-life requirements and generating industrial code exists, but is still relatively scarce \cite{koziolekLLMbasedControlCode2024, patilSpecificationDrivenLLMBasedGeneration2025, weiRequirementsAreAll2024, liuEmpiricalStudyCode2024}.

Rapid processing of large amounts of requirements in natural language is crucial for fast development cycles.
This is especially important in the automotive industry, as cars are becoming software-defined commodities and the amount of onboard code has skyrocketed in recent years.
A modern vehicle is estimated to require between 100 and 500 million lines of code \cite{madni2023handbook}.
In light of this enormous complexity, the automotive industry must undergo a paradigm shift from the classical and inflexible development strategies, such as the V cycle \cite{liuIncrementalVModelProcess2016}, to more modern approaches with shorter development cycles and quick feedback.
Several works try to incorporate generative AI, and especially LLMs, to achieve this goal \cite{shiAegisAdvancedLLMBased2024, ullrichExpandingClassicalVModel2024, katzourakisSystemsEngineeringAutonomous2025}.

This work examines how well LLMs generate simulation configuration code from high-level requirements.
We design an LLM-based system capable of quickly processing the given requirements and producing the configuration code for the CARLA simulator \cite{dosovitskiyCARLAOpenUrban2017}.
This system is supposed to aid in rapid development and prototyping, where the user can quickly modify the vehicle setup or the testing conditions by using natural language descriptions.
We tested our system by feeding it high-level requirements regarding the Autonomous Emergency Braking (AEB) System and the vehicle's sensors.

\begin{figure}[ht]
\centering
\includegraphics[width=0.8\columnwidth]{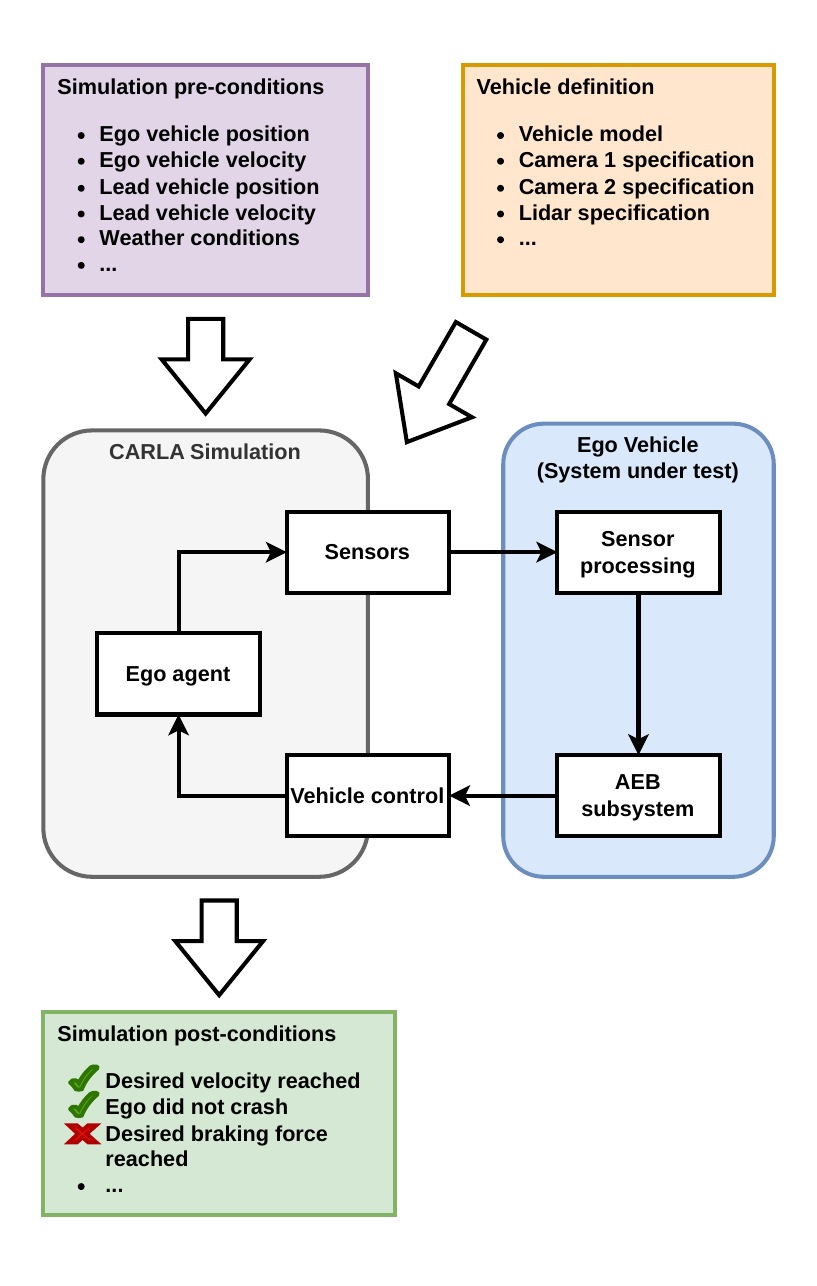}
\caption{Overview of the test setup. Simulation pre-conditions (purple) refer to the setup of the scene - agent positions, their desired behavior, weather conditions, etc. Simulation post-conditions (green) reflect the desired outcomes of the test and are largely related to the vehicle's telemetry. Vehicle definition (orange) is responsible for specifying the ego vehicle's sensor array. All modules in the system under test are considered to be fixed. The requirements visible in the figure are simplified.}
\label{fig:intro}
\end{figure}

The requirements considered in this work are split into three categories: vehicle description, test case pre-conditions, and test case post-conditions.
These requirement sets should be independent, meaning they can be processed separately.
The vehicle description primarily includes requirements related to its sensor setup.
The pre-conditions describe the test setup: agent placement, desired agent behavior, weather conditions, etc.
The post-conditions describe the test's desired outcomes, especially regarding the vehicle's telemetry.
The pre- and post-conditions are extracted from UN regulation No. 152 \cite{UN152}, which defines testing procedures for the AEB.
These requirements have a much higher level of complexity than most requirements found in other works \cite{zhangChatSceneKnowledgeEnabledSafetyCritical2024}.
The system under test consists of a sensor processing module and the AEB module, and they are considered fixed - no code is generated for these subsystems.
An overview of the test setup and the roles of the requirement groups listed above are presented in Fig. \ref{fig:intro}, the details of the generative system are presented in Fig. \ref{fig:methods_pipeline}, while example screenshots of test cases implemented in CARLA are visible in Fig. \ref{fig:carla_car} and Fig. \ref{fig:carla_pedestrian},.
This work does not cover the process of requirement extraction, nor their evaluation for completeness and correctness.

In summary, the main goals of this work are:
\begin{enumerate}
  \item Design of an LLM-based, end-to-end system that processes abstract and complex requirements in natural language, and produces configuration code for CARLA simulation.
  \item Evaluation of the system's performance on the three categories of requirements listed above, emphasizing processing requirements with a high level of abstraction and multiple constraints, and identifying category-specific challenges.
  \item Identification of shortcomings and limitations of the system, which should point in the right direction for future developments.
\end{enumerate}

\section{Background and related works}

Three main approaches exist to generate test scenarios for autonomous driving: knowledge-based, data-driven, and adversarial.
Knowledge-based methods depend on expert rules, e.g., using crowd models \cite{caiSUMMITSimulatorUrban2019} or traffic simulators with certain constraints \cite{klischatScenarioFactoryCreating2020}.
The data-driven method for scenario generation relies on real-world data \cite{tanLanguageConditionedTraffic2023}.
In adversarial approaches, adversarial agents are trained or modeled to exploit the autonomous agents \cite{yangSuicidalPedestrianGeneration2023}.
Mixed approaches include \cite{zhangChatSceneKnowledgeEnabledSafetyCritical2024, haoBridgingDataDrivenKnowledgeDriven2023}. 
Our approach, where scenarios are generated from highly standardized requirements, fits between knowledge-based and data-driven methods.

The use of LLMs in automotive test scenario generation pipelines remains relatively unexplored.
An example of such attempts includes ChatScene \cite{zhangChatSceneKnowledgeEnabledSafetyCritical2024}, where an LLM-based agent generates semi-formal descriptions of test scenarios and maps these descriptions onto Scenic code snippets to obtain a working simulation.
Another work is LCTGen \cite{tanLanguageConditionedTraffic2023}, which uses an LLM to convert a natural language specification of a test scenario into a compressed form, later used to generate the actual test.
To our knowledge, no system attempts to generate scenario definitions directly from high-level requirements, without any intermediate representation or code database.
Moreover, the scenario description is only a part of the generated code; the other part is related to vehicle definition, making our system more comprehensive.

Code generation using LLMs is a topic with ongoing and dynamic research \cite{jiang2024surveylargelanguagemodels, jin2024llmsllmbasedagentssoftware}.
Many efforts are concentrated on benchmark tasks \cite{chen2021evaluatinglargelanguagemodels, liu2023codegeneratedchatgptreally, austin2021programsynthesislargelanguage}.
Nevertheless, code generation in industrial contexts remains a challenge. \cite{liuEmpiricalStudyCode2024} examines code generation based on requirements and introduces a novel prompting technique, though it necessitates substantial preprocessing of the original requirements.
\cite{weiRequirementsAreAll2024} proposes a waterfall development process, where the LLM is asked to gradually refine the requirements and develop intermediate artifacts.
Contrary to the abovementioned works, we intend to transform requirements into code using a black-box generative system.
By evaluating the performance of our system when faced with real-life requirements, we intend to shed more light on the limitations of LLMs and possible design choices that may help alleviate them.

A task that has many parallels to generating configuration code is the generation of model instances based on metamodels.
Indeed, configuration description, like JSON schema, can be understood as metamodels, while the corresponding configuration instances are instance models.
Several works target the generation of models using LLMs - \cite{PanGenerativeAIForOCL, abukhalaf2023codexpromptengineeringocl, diroccoUseLargeLanguage2025}.

While increasing LLM model size has probably been the main factor in improving performance, the way the user interacts with the model allows it to reach its full potential.
In this spirit, multiple prompting techniques have been developed, like In-Context Learning (ICL) \cite{dongSurveyIncontextLearning2024}, where the prompt is enriched by example input-output pairs, Chain-of-Thought (CoT) \cite{weiChainThoughtPromptingElicits2022}, which further expands on the ICL prompting by adding a chain of reasoning that leads from the input to the desired output, or ReAct \cite{yaoReActSynergizingReasoning2022}, where the LLM is asked to alternate between reasoning and acting (e.g., to retrieve additional information) to solve the given task. All these prompting techniques are designed to enable (or enhance) the reasoning abilities of LLMs. 
Although the most desirable outcome is when the task at hand can be resolved using a generic prompt, we understand that the currently available LLMs may not always be capable of solving the problem independently.
Several prompting techniques are tested in the scope of this work, which allows for pinpointing the limitations of the available models.
\section{Methods}

\begin{figure*}[ht]
\centering
\includegraphics[width=\textwidth]{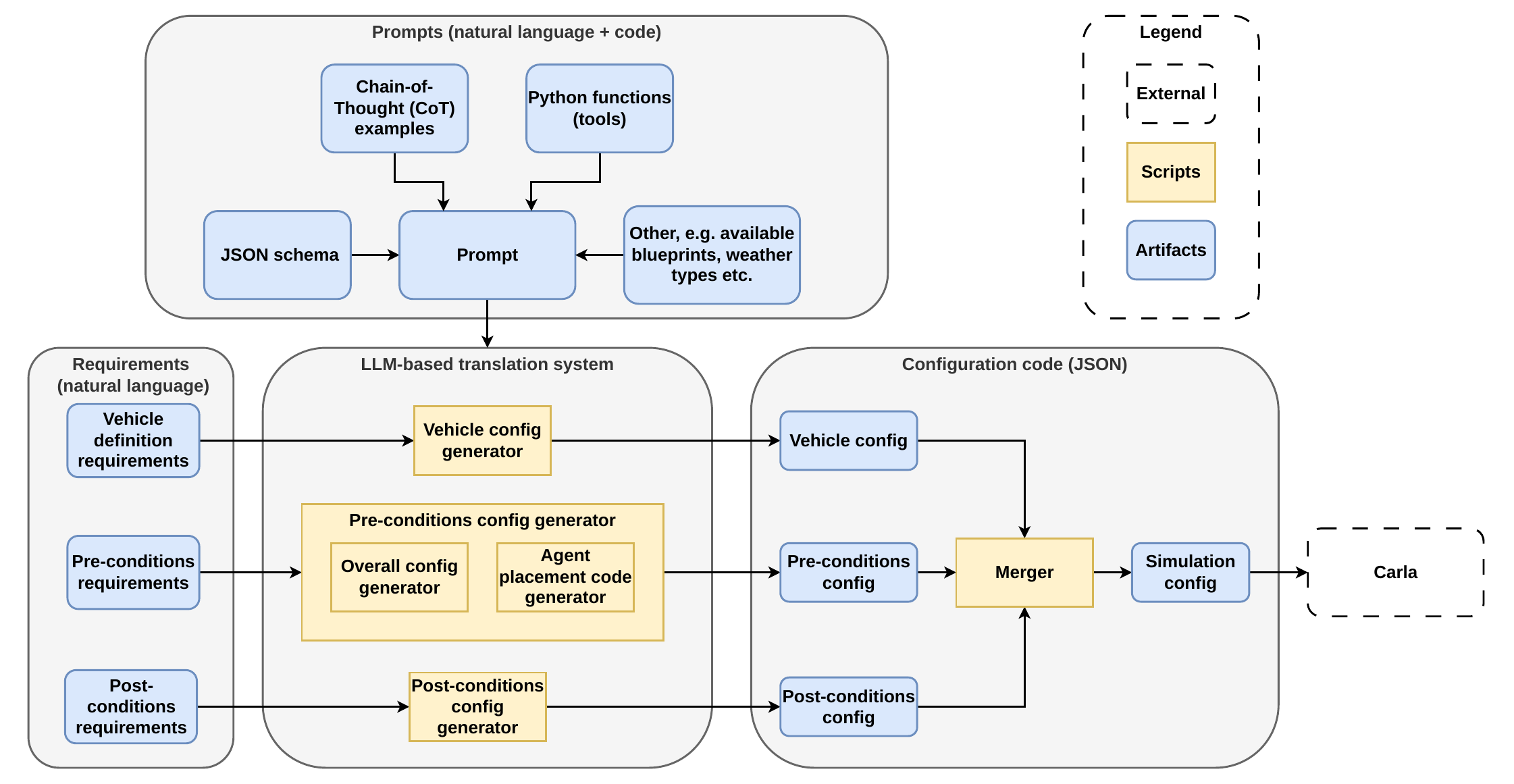}
\caption{Overview of the generative system.}
\label{fig:methods_pipeline}
\end{figure*}

In the scope of this work, we intend to generate comprehensive test scenarios based on high-level requirements in natural language.
A JSON configuration file represents a single test scenario.
The configuration consists of three parts with the following content: vehicle definition, test case pre-conditions, and test case post-conditions.
All these parts are generated separately, and later merged into a single JSON file, which can be fed into a simulation runner based on CARLA \cite{dosovitskiyCARLAOpenUrban2017}.
An overview of the pipeline is available in Fig. \ref{fig:methods_pipeline}.
As visible in the figure, there are three separate, generative, LLM-based modules.
The LLM model used for evaluation is GPT-4o.

\subsection{Requirement types}
\label{section:methods_req_types}

After analysis of the requirements from the available datasets, we categorized them into three categories (listed below).
This categorization may help understand specific properties of the generative systems presented in our work.

\begin{enumerate}
    \item \emph{Direct} - map directly to available configuration parameters.
    \item \emph{Indirect} - map to multiple configuration parameters or map directly, but after a certain transformation. These requirements do not require external context; the LLM's internal knowledge should be enough to map them properly. Examples would be a translation of camera resolution expressed in megapixels into image width and height, or the transformation of frequency into period.
    \item \emph{Abstract} - additional knowledge (context) is required to map them properly onto configuration parameters. Context can be provided by other requirements (e.g., bounding box of the vehicle for sensor placement) or external means (e.g., a graph of road network for agent placement).
\end{enumerate}

\subsection{Vehicle definition pipeline}

The requirements for vehicle definitions mainly contain sensor specifications.
Appendix \ref{appendix:vehicle_def_req} shows an example subset of requirements for this pipeline.
The \emph{Vehicle Configuration Generator} generates the configuration files in a single step.
In addition to the requirements set, the prompts are parametrized with the relevant JSON schema, which describes the desired structure of the resulting configuration file and available CARLA blueprints.

\subsection{Pre-conditions pipeline}

The sets of requirements for the pre-conditions are extracted from UN regulation No. 152 \cite{UN152}, with examples presented in Appendix \ref{appendix:precondition_req}.
After analysis of the requirements for the pre-conditions, we decided to extend the \emph{Pre-condition Configuration Generator} to work in a two-step manner.
First, the overall configuration structure is generated, which contains all parameters that can be extracted directly from the requirement set.
In the second step, we ask the LLM to write a Python function that will be used to fill in the remaining parameters.
The function must have a specific signature because the system executes it automatically.
This split was necessary because translating \emph{abstract} requirements into configuration parameters requires external context.
Examples of such properties are: agent position on the road graph, their global rotation, etc.
The agents' relative position is complicated and is expressed in a convoluted way across multiple requirements.
The LLM can access several tools (Python functions) for graph handling and transforming various metrics (e.g., Time-to-Collision to distance).
In addition to the requirements set, the prompts are parametrized with the relevant JSON schema, which describes the desired structure of the resulting configuration file and available CARLA weather types.

\subsection{Post-conditions pipeline}

The requirement sets for the test post-conditions are extracted from UN regulation No. 152 \cite{UN152}, with an example dataset presented in Appendix \ref{appendix:postcondition_req}.
The \emph{Post-Condition Configuration Generator} generates the configuration files in a single step.
These requirements relate primarily to the ego vehicle's telemetry - speed, braking force, etc.
In addition to the requirement set, the prompts are parametrized with the relevant JSON schema, which describes the desired structure of the resulting configuration file, available telemetry options, and a set of pre-defined events, which can be used to perform relevant telemetry checks.
Event examples are the start and end of the simulation and the beginning and end of the AEB braking.
Telemetry checks can be performed at a single time point or in a time range.
The available events are currently pre-defined, but we intend to automatically extract them from the requirements in the future.

\subsection{Merger}

The merging step resolves discrepancies when processing requirements from multiple sources and aligning them with what CARLA needs.
An example is the different name of the vehicle under test - \emph{ego} in the vehicle definition requirements, and \emph{subject} in the pre- and post-condition requirements, or injection of a collision sensor, which is usually not defined as part of the requirements, but which is helpful for assessment of the results.

\subsection{Prompt types}
\label{section:methods_prompt_types}

Three prompting styles are put to the test:
\begin{enumerate}
  \item \emph{Simple} - A generic prompt with no references to the requirements or configuration structure, parametrized with the requirements and the relevant JSON schema
  \item \emph{In-Context Learning (ICL)} - The Simple prompt extended with an example pair of requirements and the corresponding JSON configuration
  \item \emph{Chain of Thought (CoT)} - The ICL prompt extended with reasoning examples that allow for translating the most problematic requirements into code.
\end{enumerate}

\subsection{Evaluation methods}
\label{section:methods_evaluation}

The quality of the resulting configuration code is evaluated using assertions, which are similar to unit tests.
The following metrics are used to evaluate the performance of the system:
\begin{enumerate}
    \item \emph{Average Test Passing Rate (TPR)} - ratio of the number of passed assertions (tests) to the total number of tests, averaged over multiple code generation attempts.
    \item \emph{Pass@k} – statistical metric, which tells how likely it is to find at least one correct (TPR = 1) configuration in a randomly selected set of configurations of size k.
\end{enumerate}

\section{Experiments and result analysis}

The metrics used in the following sections are explained in \ref{section:methods_evaluation}, the prompting styles are described in \ref{section:methods_prompt_types}, and the requirement types are described in \ref{section:methods_req_types}.

\subsection{Vehicle definition - automatic generation of sensor configuration}
\label{sec:vehicle_def_1}

\begin{figure}[ht]
\centering
\begin{subfigure}
    \centering
    \includegraphics[width=0.45\textwidth]{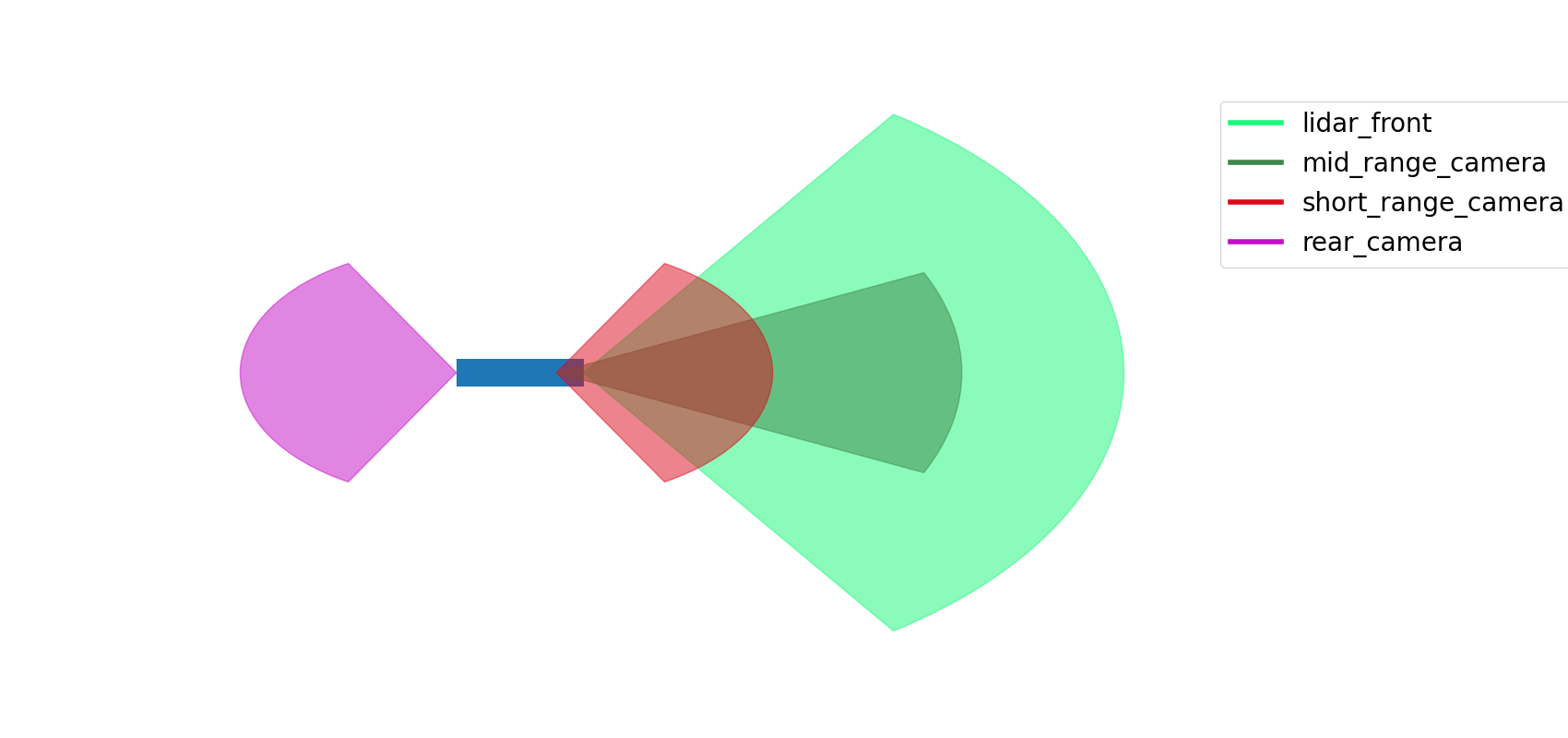}
    \caption{Visualization of an example dataset for the vehicle definition pipeline. 4 sensors have been defined in the requirements.}
    \label{fig:vehicle_def_sensors}
\end{subfigure}
\par\bigskip
\begin{subfigure}
    \centering
    \includegraphics[width=0.45\textwidth]{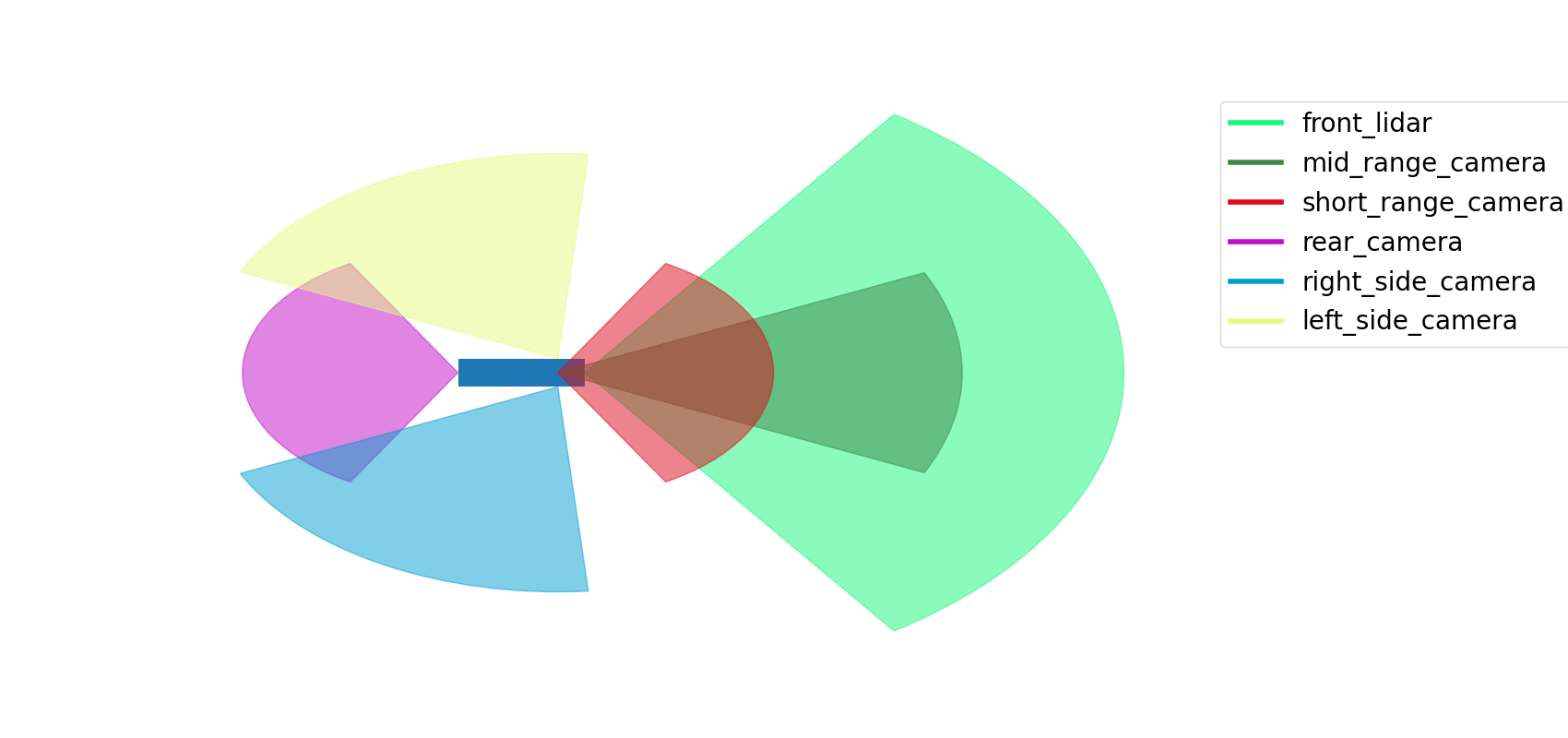}
    \caption{The requirements from Fig. \ref{fig:vehicle_def_sensors} have been extended with 2 additional sensors. The system is able to generate the relevant configuration directly from the requirements.}
    \label{fig:vehicle_def_sensors_extended}
\end{subfigure}
\end{figure}

Fig. \ref{fig:vehicle_def_sensors} shows a visualization of the vehicle's sensor configurations generated automatically from example requirements in natural language.
When the requirements are extended with two additional sensors (Fig. \ref{fig:vehicle_def_sensors_extended}), the system regenerates the configuration, taking into account the sensors' position on the vehicle, their rotation, field of view, etc.
When passed to the CARLA runner, all sensors defined in the configuration will be spawned automatically.
The two datasets (base and extended) are presented in Appendix \ref{appendix:vehicle_def_basic}, together with the LLM-generated configuration code.

\subsection{Vehicle definition - prompting style}
\label{sec:vehicle_def_2}

This experiment explores how various prompting techniques affect the performance of the vehicle definition pipeline.
The dataset contains 33 requirements, which define a vehicle with a sensor setup similar to that of a Tesla.
Table \ref{table:vehicle_def_prompt} presents the results of 50 passes for each prompt type - \emph{Simple}, \emph{In-Context-Learning (ICL)}, and \emph{Chain-of-Thought (CoT)}.
The \emph{CoT} prompt was extended with a single example of reasoning for each of the two requirements that had the highest rate of failures with the \emph{Simple} and \emph{ICL} prompts.
The results show that the LLM has difficulties in translating \emph{indirect} and \emph{abstract} requirements into code without additional help.
The most challenging requirements to translate are those concerning sensor placement.
The placement is specified relative to the vehicle’s bounding box.
Only with the added examples of reasoning is the system able to process them correctly.

\begin{table}[!h]
\caption{Impact of prompting techniques on the vehicle definition pipeline.}
\label{table:vehicle_def_prompt}
\begin{center}
\begin{tabular}{ |c| c c c| }
\hline
Metric & \multicolumn{3}{c|}{Prompt type} \\
       & Simple & ICL & COT \\
\hline
Avg TPR & 0.81 & 0.86 & 0.99 \\  
Pass@1  & 0.0 & 0.0 & 0.80 \\
Pass@5  & 0.0 & 0.0 & 1.0 \\
Pass@10 & 0.0 & 0.0 & 1.0 \\
Pass@20 & 0.0 & 0.0 & 1.0\\
 \hline
\end{tabular}
\end{center}
\end{table}

\subsection{Vehicle definition - requirement order}
\label{sec:vehicle_def_3}

The third experiment explores how the requirement order impacts the performance of the vehicle definition pipeline.
Table \ref{table:vehicle_def_shuffle} presents the impact of requirement shuffling on the vehicle definition pipeline.
The \emph{ordered} column uses a dataset where the requirements are logically grouped and ordered, e.g., all requirements pertinent to a single sensor are put one after another.
The \emph{shuffled} column uses a dataset where the requirements were shuffled.
The prompt type was set to \emph{CoT} for this experiment, with the same reasoning examples as in the experiment presented in section \ref{sec:vehicle_def_2}.
The results show that logical grouping of requirements is an important factor that significantly affects the system performance.

\begin{table}[!h]
\caption{Impact of requirement shuffling on the vehicle definition pipeline.}
\label{table:vehicle_def_shuffle}
\begin{center}
\begin{tabular}{ |c|cc| }
\hline
Metric & \multicolumn{2}{c|}{Requirement order} \\
       & Ordered & Shuffled \\
\hline
Avg TPR & 0.99 & 0.89 \\  
Pass@1  & 0.8  & 0.1  \\
Pass@5  & 1.0  & 0.42 \\
Pass@10 & 1.0  & 0.69 \\
Pass@20 & 1.0  & 0.93 \\
 \hline
\end{tabular}
\end{center}
\end{table}

\subsection{Vehicle definition - performance as function of number of requirements}

\begin{figure}[ht]
\centering
\begin{subfigure}
    \centering
    \includegraphics[width=0.4\textwidth]{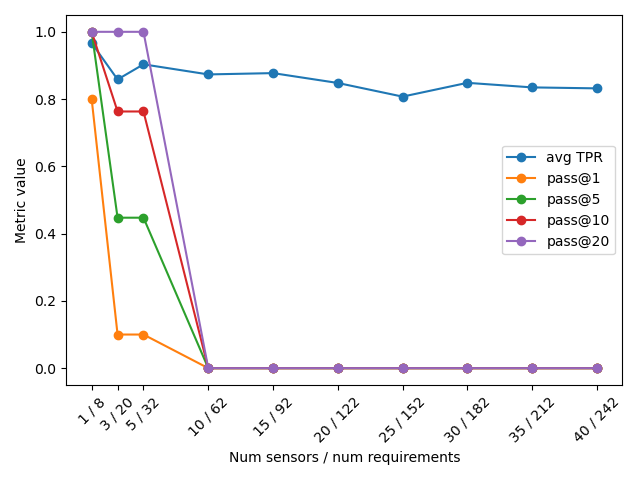}
    \caption{Performance of the system as a function of the number of requirements (no requirement grouping)}
    \label{fig:vehicle_def_no_grouping}
\end{subfigure}
\begin{subfigure}
    \centering
    \includegraphics[width=0.4\textwidth]{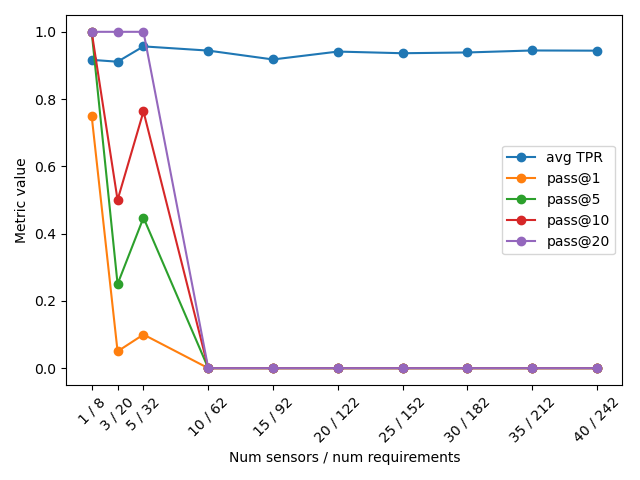}
    \caption{Performance of the system as a function of the number of requirements (with requirement grouping)}
    \label{fig:vehicle_def_grouping}
\end{subfigure}
\end{figure}

Fig. \ref{fig:vehicle_def_no_grouping} and Fig. \ref{fig:vehicle_def_grouping} present the results of the analysis of how the performance of the vehicle definition pipeline depends on the size of the requirements set.
The datasets for the experiment are generated programmatically from a parameterized set of requirements that describe a camera sensor.
This parameterized set contains 3 \emph{direct} requirements, 2 \emph{indirect} ones, and a single \emph{abstract} requirement.

The results visible in Fig. \ref{fig:vehicle_def_no_grouping} show that the values of the \emph{pass@k} metrics drop pretty drastically.
However, the \emph{TPR} metric suggests that the system is still able to process the majority of the requirements correctly.
Indeed, a closer look at the results shows that the system is able to process the \emph{direct} requirements very well, even when the number of requirements grows.
However, as the number of requirements increases, the system quickly loses the ability to process the \emph{indirect} and \emph{abstract} requirements.

Fig. \ref{fig:vehicle_def_grouping} shows the performance of a modified version of the system.
The requirements set is split by the LLM into logical groups.
Each of the groups is then processed separately.
Of course, this incurs a much higher cost of processing, because the number of API calls equals the number of defined sensors plus the initial split.
The results show that this version of the system is able to maintain a very high performance, as indicated by the \emph{TPR} metric, which does not decrease as the number of requirements grows.
However, the \emph{pass@k} metrics did not improve much.
This is caused by the relatively high error rate in processing the \emph{abstract} requirements.

\subsection{Simulation pre-conditions - prompting style}
\label{sec:pre_conditions1}

\begin{figure}[ht]
\centering
\begin{subfigure}
    \centering
    \includegraphics[width=0.4\textwidth]{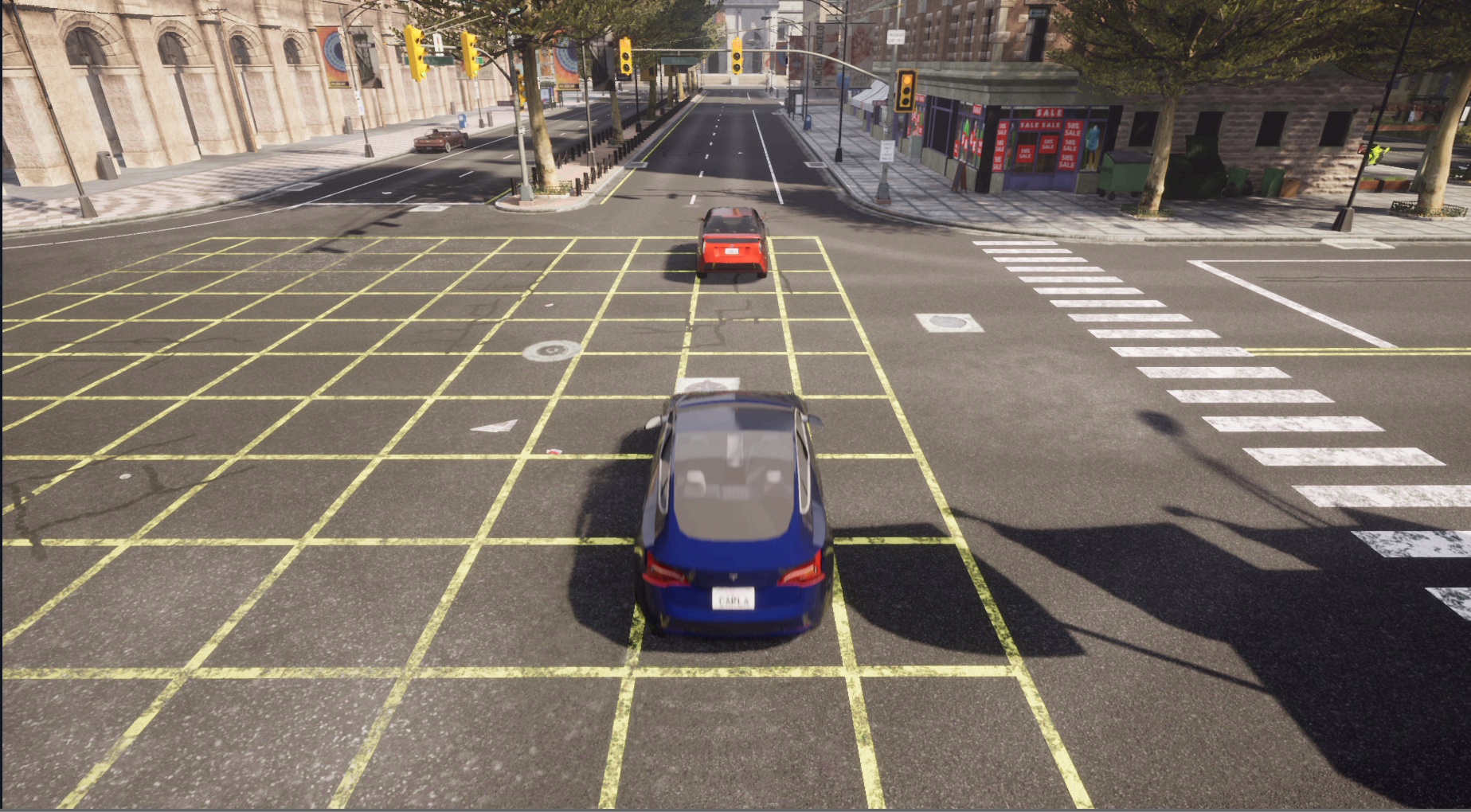}
    \caption{An example test scenario for the AEB system, with the ego vehicle following the (slower) lead vehicle.}
    \label{fig:carla_car}
\end{subfigure}
\par\bigskip
\begin{subfigure}
    \centering
    \includegraphics[width=0.4\textwidth]{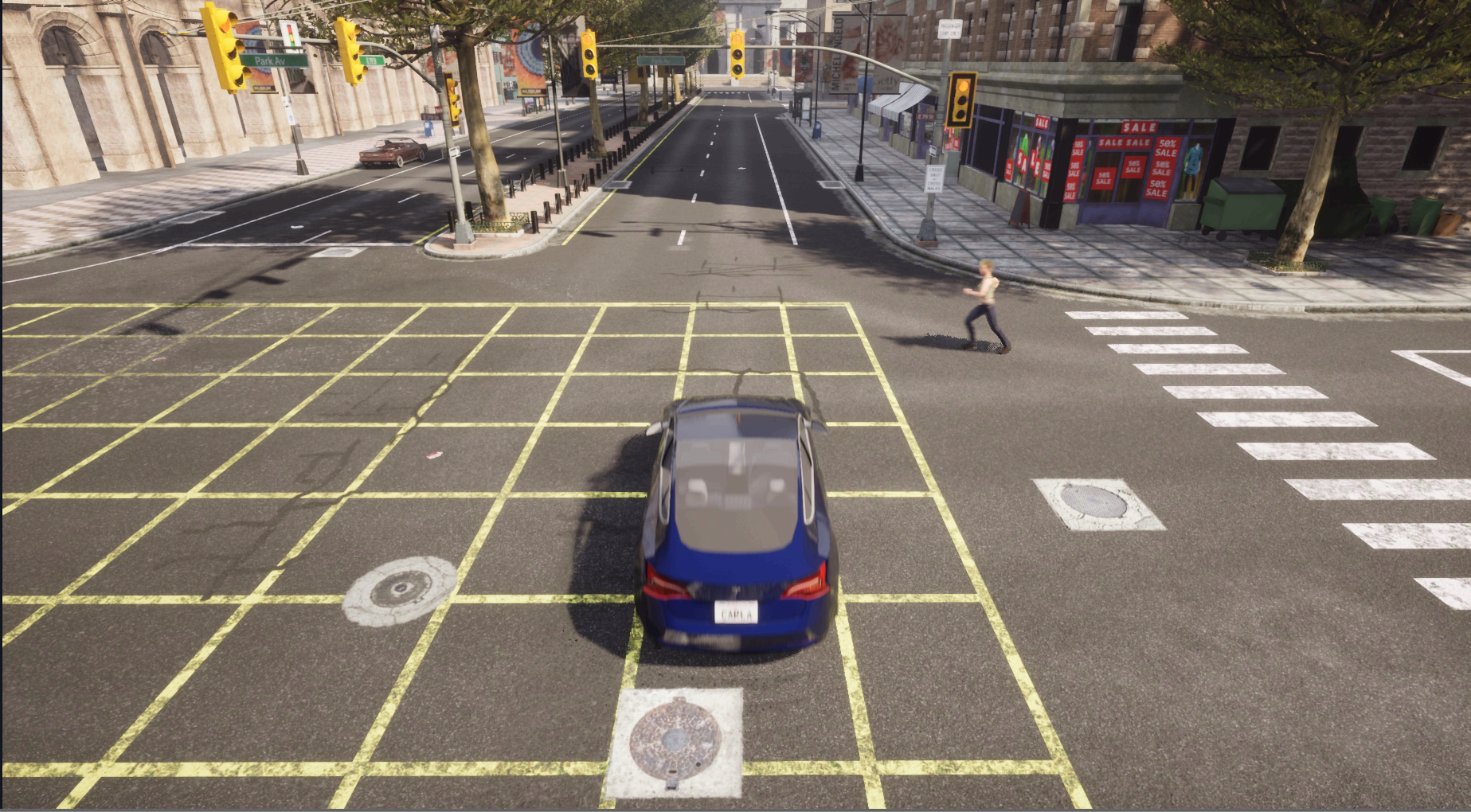}
    \caption{An example test scenario for the AEB system, with a pedestrian crossing the path of the ego vehicle. The trajectories of the two agents have to be properly synchronized.}
    \label{fig:carla_pedestrian}
\end{subfigure}
\end{figure}

In this experiment, the impact of prompting type on the performance of the pre-conditions pipeline is explored.
The datasets define the pre-conditions of tests for the AEB system, and are extracted from the UN regulation No. 152 (examples in Appendix \ref{appendix:precondition_req}).
Generally, the test cases extracted from the standard can be categorized either as car-to-car or car-to-pedestrian scenarios.
In the former (Fig. \ref{fig:carla_car}), the subject vehicle follows a slower lead vehicle, and is supposed to stop before it hits the lead.
In the latter (Fig. \ref{fig:carla_pedestrian}), the subject vehicle is expected to stop in front of a pedestrian who traverses the road.
The main issue during the processing of the requirements for this pipeline is the dependency on the available road graph, as well as a very convoluted description of the agent placement.
The distance between the ego and lead vehicle requires processing several requirements, as well as translating a temporal metric (Time-to-Collision, TTC) into actual distance.
The situation is even more complicated in car-to-pedestrian scenarios, where the subject vehicle and the pedestrian must be synchronized to reach the target test point at the right time.
Given the above, the main challenge of this pipeline is to write Python code for agent placement.
The LLM has access to a number of Python functions (tools), that allows to calculate the distances between agents, extract relevant road segments from the road graph, or create CARLA spawnpoints.
An example of generated code is presented in Appendix \ref{appendix:precondition_req}.
Table \ref{table:preconditions_car_to_car_prompt} presents the results of 20 passes on an example car-to-car scenario with the following prompt types - \emph{Simple}, and \emph{ICL / CoT}.
Table \ref{table:preconditions_car_to_pedestrian_prompt} presents the results of 20 passes on an example car-to-pedestrian scenario with the following prompt types - \emph{Simple}, and \emph{ICL / CoT}.
The \emph{ICL} and \emph{CoT} prompts are put together because Python code examples can be understood as Chain-of-Though examples, and the distinction between the two prompting types is blurry.
The \emph{ICL / CoT} prompt was extended with a single example of similar requirements and the corresponding Python code.
An additional metric (Code gen. success rate) is added in table \ref{table:preconditions_car_to_pedestrian_prompt}, which shows the success rate of valid Python code generation.

\begin{table}[!h]
\caption{Impact of prompting techniques on the pre-conditions pipeline (car-to-car scenario).}
\label{table:preconditions_car_to_car_prompt}
\begin{center}
\begin{tabular}{ |c|cc| }
\hline
Metric & \multicolumn{2}{c|}{Prompt type} \\
       & Simple & ICL / COT \\
 \hline
Avg TPR & 0.78 & 0.95 \\  
Pass@1  & 0.0  & 0.65 \\
Pass@5  & 0.0  & 1.0  \\
Pass@10 & 0.0  & 1.0  \\
 \hline
\end{tabular}
\end{center}
\end{table}

\begin{table}[!h]
\caption{Impact of prompting techniques on the pre-conditions pipeline (car-to-pedestrian scenario).}
\label{table:preconditions_car_to_pedestrian_prompt}
\begin{center}
\begin{tabular}{ |c|cc| }
\hline
Metric & \multicolumn{2}{c|}{Prompt type} \\
       & Simple & ICL / COT \\
 \hline
Avg TPR & 0.59 & 0.7 \\  
Code gen. success rate & 0.0 & 0.7 \\
Pass@1  & 0.0  & 0.0 \\
Pass@5  & 0.0  & 0.0  \\
Pass@10 & 0.0  & 0.0  \\
 \hline
\end{tabular}
\end{center}
\end{table}

A closer look at the results in the car-to-car scenario shows that the LLM most often fails at correctly calculating the distance between the vehicles.
Requirement 3 from the dataset seems to be particularly difficult to understand by the LLM, and the minimal distance that corresponds to the required interval of 2 seconds is often missing from the calculations.
Another issue is that there is an implied "speed-up" time that is necessary for the ego vehicle to reach the desired speed.
This is another term that is often missing from the distance calculations.

Results for the car-to-pedestrian scenario show that this case is difficult for the LLM to solve, even when a \emph{CoT} example is provided.
Without the example, the LLM is unable to generate valid code, and only some of the configuration parameters are set.
With \emph{CoT} examples, the LLM is able to generate valid Python code most of the time, but as indicated by the pass@k metrics, it is still unable to grasp all the details necessary to calculate agent positions, the trigger (synchronization) point for the pedestrian, etc.

\subsection{Simulation post-conditions - prompting style}
\label{sec:post_conditions1}

This experiment investigates how different types of prompting affect the performance of the post-condition pipeline.
The datasets, derived from UN Regulation No. 152, specify the post-conditions of tests for the AEB system (see examples in Appendix \ref{appendix:postcondition_req}).
Table \ref{table:postconditions_prompt} shows the results from 20 runs for each prompt type.
The \emph{CoT} prompt was enhanced with a single example illustrating the translation of a requirement into a telemetry check between two events.

\begin{table}[!h]
\caption{Impact of prompting techniques on the post-conditions pipeline.}
\label{table:postconditions_prompt}
\begin{center}
\begin{tabular}{ |c| c c c| }
\hline
Metric & \multicolumn{3}{c|}{Prompt type} \\
       & Simple & ICL & COT \\
 \hline
Avg TPR & 0.84 & 0.98 & 1.0 \\  
Pass@1  & 0.45 & 0.9  & 1.0 \\
Pass@5  & 0.97 & 1.0  & 1.0 \\
Pass@10 & 1.0  & 1.0  & 1.0 \\
 \hline
\end{tabular}
\end{center}
\end{table}

This pipeline tests the LLM’s capability to understand the temporal sequence of events.
While the dataset is relatively small, the pipeline demonstrates promising performance.
\section{Conclusions}

This work introduces an end-to-end, LLM-based system that transforms abstract natural language requirements into configuration code for automotive simulations.
A key advantage of this approach is the reduced need for users to understand the internal mechanics and parameters of the configuration system.
As a result, users without specialized knowledge of the simulation environment can interact with it solely through natural language.

The experiments show that the LLMs are a valuable tool for rapid requirement translation, and that they can be used to translate \emph{direct} requirements with a high level of confidence.
However, they still struggle with \emph{indirect}, especially with \emph{abstract} requirements, even when providing Chain-of-Thought examples.
The requirements extracted from the UN regulation have a much higher level of complexity than requirements used in other works \cite{zhangChatSceneKnowledgeEnabledSafetyCritical2024}.
For instance, the minimal distance between the ego and the lead vehicles is defined by multiple requirements in a convoluted way (see Appendix \ref{appendix:precondition_req}).

Although our setup targets testing the AEB system, we believe it could easily be adapted to other automotive systems and standards.
An example would be the UN Regulation No. 157 - Automated Lane Keeping Systems (ALKS) \cite{UN157}.
The one thing that may be extended is the repertoire of available Python tools for agent placement.

For the \emph{abstract} requirements present in our datasets for scenario definition, using a more expressive language, like Scenic \cite{fremontScenicLanguageScenario2023}, could be beneficial.
This, however, comes with difficulties of its own.
Being a Domain Specific Language (DSL), Scenic is likely to be problematic to generate for currently available LLMs without additional tuning \cite{joelSurveyLLMbasedCode2024}.
We are also investigating the possibility of using Scenic to assert the telemetry based on the post-conditions of the test.

The datasets used in this work were extracted manually from the relevant standards.
However, as the next step, we intend to use an automated, LLM-based RAG system, similar to \cite{zolfaghariAdoptingRAGLLMAided2024} for test case extraction.
Combining these two systems will create a comprehensive pipeline for automotive system testing, from documents and standards to a running simulation.

Finally, to enable generation of code that would reflect the behavioral aspects of the desired assisted driving functionality (such as emergency braking and assisted parking), we plan to integrate the work presented in this paper with \cite{pan2025automatingautomotivesoftwaredevelopment}, where we introduce event chain-driven code generation leveraging model-driven engineering in synergy with LLMs.

\bibliographystyle{IEEEtran}
\bibliography{IEEEabrv, bibliography}

\begin{thebibliography}{10}
\providecommand{\url}[1]{#1}
\csname url@samestyle\endcsname
\providecommand{\newblock}{\relax}
\providecommand{\bibinfo}[2]{#2}
\providecommand{\BIBentrySTDinterwordspacing}{\spaceskip=0pt\relax}
\providecommand{\BIBentryALTinterwordstretchfactor}{4}
\providecommand{\BIBentryALTinterwordspacing}{\spaceskip=\fontdimen2\font plus
\BIBentryALTinterwordstretchfactor\fontdimen3\font minus \fontdimen4\font\relax}
\providecommand{\BIBforeignlanguage}[2]{{%
\expandafter\ifx\csname l@#1\endcsname\relax
\typeout{** WARNING: IEEEtran.bst: No hyphenation pattern has been}%
\typeout{** loaded for the language `#1'. Using the pattern for}%
\typeout{** the default language instead.}%
\else
\language=\csname l@#1\endcsname
\fi
#2}}
\providecommand{\BIBdecl}{\relax}
\BIBdecl

\bibitem{HealthAI}
\BIBentryALTinterwordspacing
S.~Goyal, E.~Rastogi, S.~P. Rajagopal, D.~Yuan, F.~Zhao, J.~Chintagunta, G.~Naik, and J.~Ward, ``Healai: A healthcare llm for effective medical documentation,'' in \emph{Proceedings of the 17th ACM International Conference on Web Search and Data Mining}, ser. WSDM '24.\hskip 1em plus 0.5em minus 0.4em\relax New York, NY, USA: Association for Computing Machinery, 2024, p. 1167–1168. [Online]. Available: \url{https://doi.org/10.1145/3616855.3635739}
\BIBentrySTDinterwordspacing

\bibitem{cascellaEvaluatingFeasibilityChatGPT2023}
M.~Cascella, J.~Montomoli, V.~Bellini, and E.~Bignami, ``Evaluating the {{Feasibility}} of {{ChatGPT}} in {{Healthcare}}: {{An Analysis}} of {{Multiple Clinical}} and {{Research Scenarios}},'' \emph{Journal of Medical Systems}, vol.~47, no.~1, p.~33, Mar. 2023.

\bibitem{zhaoRevolutionizingFinanceLLMs2024}
H.~Zhao, Z.~Liu, Z.~Wu, Y.~Li, T.~Yang, P.~Shu, S.~Xu, H.~Dai, L.~Zhao, H.~Jiang, Y.~Pan, J.~Chen, Y.~Zhou, G.~Mai, N.~Liu, and T.~Liu, ``Revolutionizing {{Finance}} with {{LLMs}}: {{An Overview}} of {{Applications}} and {{Insights}},'' 2024.

\bibitem{radivojevicHumanPerceptionLLMgenerated2024}
K.~Radivojevic, M.~Chou, K.~{Badillo-Urquiola}, and P.~Brenner, ``Human {{Perception}} of {{LLM-generated Text Content}} in {{Social Media Environments}},'' 2024.

\bibitem{yangSocialMindLLMbasedProactive2024}
B.~Yang, Y.~Guo, L.~Xu, Z.~Yan, H.~Chen, G.~Xing, and X.~Jiang, ``{{SocialMind}}: {{LLM-based Proactive AR Social Assistive System}} with {{Human-like Perception}} for {{In-situ Live Interactions}},'' 2024.

\bibitem{paulBenchmarksMetricsEvaluations2024}
D.~G. Paul, H.~Zhu, and I.~Bayley, ``Benchmarks and {{Metrics}} for {{Evaluations}} of {{Code Generation}}: {{A Critical Review}},'' Jun. 2024.

\bibitem{koziolekLLMbasedControlCode2024}
H.~Koziolek and A.~Koziolek, ``{{LLM-based Control Code Generation}} using {{Image Recognition}},'' in \emph{Proceedings of the 1st {{International Workshop}} on {{Large Language Models}} for {{Code}}}.\hskip 1em plus 0.5em minus 0.4em\relax Lisbon Portugal: ACM, Apr. 2024, pp. 38--45.

\bibitem{patilSpecificationDrivenLLMBasedGeneration2025}
M.~S. Patil, G.~Ung, and M.~Nyberg, ``Towards {{Specification-Driven LLM-Based Generation}} of {{Embedded Automotive Software}},'' in \emph{Bridging the {{Gap Between AI}} and {{Reality}}}, B.~Steffen, Ed.\hskip 1em plus 0.5em minus 0.4em\relax Cham: Springer Nature Switzerland, 2025, vol. 15217, pp. 125--144.

\bibitem{weiRequirementsAreAll2024}
B.~Wei, ``Requirements are {{All You Need}}: {{From Requirements}} to {{Code}} with {{LLMs}},'' in \emph{2024 {{IEEE}} 32nd {{International Requirements Engineering Conference}} ({{RE}})}.\hskip 1em plus 0.5em minus 0.4em\relax Reykjavik, Iceland: IEEE, Jun. 2024, pp. 416--422.

\bibitem{liuEmpiricalStudyCode2024}
M.~Liu, J.~Wang, T.~Lin, Q.~Ma, Z.~Fang, and Y.~Wu, ``An {{Empirical Study}} of the {{Code Generation}} of {{Safety-Critical Software Using LLMs}},'' \emph{Applied Sciences}, vol.~14, no.~3, p. 1046, Jan. 2024.

\bibitem{madni2023handbook}
\BIBentryALTinterwordspacing
A.~Madni, N.~Augustine, and M.~Sievers, \emph{Handbook of Model-Based Systems Engineering}, ser. Handbook of Model-Based Systems Engineering.\hskip 1em plus 0.5em minus 0.4em\relax Springer International Publishing, 2023. [Online]. Available: \url{https://books.google.de/books?id=CFLNEAAAQBAJ}
\BIBentrySTDinterwordspacing

\bibitem{liuIncrementalVModelProcess2016}
B.~Liu, H.~Zhang, and S.~Zhu, ``An {{Incremental V-Model Process}} for {{Automotive Development}},'' in \emph{2016 23rd {{Asia-Pacific Software Engineering Conference}} ({{APSEC}})}.\hskip 1em plus 0.5em minus 0.4em\relax Hamilton, New Zealand: IEEE, 2016, pp. 225--232.

\bibitem{shiAegisAdvancedLLMBased2024}
L.~Shi, B.~Qi, J.~Luo, Y.~Zhang, Z.~Liang, Z.~Gao, W.~Deng, and L.~Sun, ``Aegis:{{An Advanced LLM-Based Multi-Agent}} for {{Intelligent Functional Safety Engineering}},'' 2024.

\bibitem{ullrichExpandingClassicalVModel2024}
L.~Ullrich, M.~Buchholz, K.~Dietmayer, and K.~Graichen, ``Expanding the {{Classical V-Model}} for the {{Development}} of {{Complex Systems Incorporating AI}},'' \emph{IEEE Transactions on Intelligent Vehicles}, pp. 1--15, 2024.

\bibitem{katzourakisSystemsEngineeringAutonomous2025}
D.~Katzourakis, ``Systems {{Engineering}} for {{Autonomous Vehicles}}; {{Supervising AI}} using {{Large Language Models}} ({{SSuperLLM}}),'' 2025.

\bibitem{dosovitskiyCARLAOpenUrban2017}
A.~Dosovitskiy, G.~Ros, F.~Codevilla, A.~Lopez, and V.~Koltun, ``{{CARLA}}: {{An Open Urban Driving Simulator}},'' 2017.

\bibitem{UN152}
``Un regulation no 152 – uniform provisions concerning the approval of motor vehicles with regard to the advanced emergency braking system (aebs) for m1 and n1 vehicles [2020/1597],'' pp. 66--89, Oct 2020.

\bibitem{zhangChatSceneKnowledgeEnabledSafetyCritical2024}
J.~Zhang, C.~Xu, and B.~Li, ``{{ChatScene}}: {{Knowledge-Enabled Safety-Critical Scenario Generation}} for {{Autonomous Vehicles}},'' in \emph{2024 {{IEEE}}/{{CVF Conference}} on {{Computer Vision}} and {{Pattern Recognition}} ({{CVPR}})}.\hskip 1em plus 0.5em minus 0.4em\relax Seattle, WA, USA: IEEE, Jun. 2024, pp. 15\,459--15\,469.

\bibitem{caiSUMMITSimulatorUrban2019}
P.~Cai, Y.~Lee, Y.~Luo, and D.~Hsu, ``{{SUMMIT}}: {{A Simulator}} for {{Urban Driving}} in {{Massive Mixed Traffic}},'' 2019.

\bibitem{klischatScenarioFactoryCreating2020}
M.~Klischat, E.~I. Liu, F.~Holtke, and M.~Althoff, ``Scenario {{Factory}}: {{Creating Safety-Critical Traffic Scenarios}} for {{Automated Vehicles}},'' in \emph{2020 {{IEEE}} 23rd {{International Conference}} on {{Intelligent Transportation Systems}} ({{ITSC}})}.\hskip 1em plus 0.5em minus 0.4em\relax Rhodes, Greece: IEEE, Sep. 2020, pp. 1--7.

\bibitem{tanLanguageConditionedTraffic2023}
S.~Tan, B.~Ivanovic, X.~Weng, M.~Pavone, and P.~Kraehenbuehl, ``Language {{Conditioned Traffic Generation}},'' 2023.

\bibitem{yangSuicidalPedestrianGeneration2023}
Y.~Yang, K.~Kujanpaa, A.~Babadi, J.~Pajarinen, and A.~Ilin, ``Suicidal {{Pedestrian}}: {{Generation}} of {{Safety-Critical Scenarios}} for {{Autonomous Vehicles}},'' 2023.

\bibitem{haoBridgingDataDrivenKnowledgeDriven2023}
K.~Hao, L.~Liu, W.~Cui, J.~Zhang, S.~Yan, Y.~Pan, and Z.~Yang, ``Bridging {{Data-Driven}} and {{Knowledge-Driven Approaches}} for {{Safety-Critical Scenario Generation}} in {{Automated Vehicle Validation}},'' 2023.

\bibitem{jiang2024surveylargelanguagemodels}
\BIBentryALTinterwordspacing
J.~Jiang, F.~Wang, J.~Shen, S.~Kim, and S.~Kim, ``A survey on large language models for code generation,'' 2024. [Online]. Available: \url{https://arxiv.org/abs/2406.00515}
\BIBentrySTDinterwordspacing

\bibitem{jin2024llmsllmbasedagentssoftware}
\BIBentryALTinterwordspacing
H.~Jin, L.~Huang, H.~Cai, J.~Yan, B.~Li, and H.~Chen, ``From llms to llm-based agents for software engineering: A survey of current, challenges and future,'' 2024. [Online]. Available: \url{https://arxiv.org/abs/2408.02479}
\BIBentrySTDinterwordspacing

\bibitem{chen2021evaluatinglargelanguagemodels}
\BIBentryALTinterwordspacing
M.~Chen, J.~Tworek, H.~Jun, Q.~Yuan, H.~P. de~Oliveira~Pinto, J.~Kaplan, H.~Edwards, Y.~Burda, N.~Joseph, G.~Brockman, A.~Ray, R.~Puri, G.~Krueger, M.~Petrov, H.~Khlaaf, G.~Sastry, P.~Mishkin, B.~Chan, S.~Gray, N.~Ryder, M.~Pavlov, A.~Power, L.~Kaiser, M.~Bavarian, C.~Winter, P.~Tillet, F.~P. Such, D.~Cummings, M.~Plappert, F.~Chantzis, E.~Barnes, A.~Herbert-Voss, W.~H. Guss, A.~Nichol, A.~Paino, N.~Tezak, J.~Tang, I.~Babuschkin, S.~Balaji, S.~Jain, W.~Saunders, C.~Hesse, A.~N. Carr, J.~Leike, J.~Achiam, V.~Misra, E.~Morikawa, A.~Radford, M.~Knight, M.~Brundage, M.~Murati, K.~Mayer, P.~Welinder, B.~McGrew, D.~Amodei, S.~McCandlish, I.~Sutskever, and W.~Zaremba, ``Evaluating large language models trained on code,'' 2021. [Online]. Available: \url{https://arxiv.org/abs/2107.03374}
\BIBentrySTDinterwordspacing

\bibitem{liu2023codegeneratedchatgptreally}
\BIBentryALTinterwordspacing
J.~Liu, C.~S. Xia, Y.~Wang, and L.~Zhang, ``Is your code generated by chatgpt really correct? rigorous evaluation of large language models for code generation,'' 2023. [Online]. Available: \url{https://arxiv.org/abs/2305.01210}
\BIBentrySTDinterwordspacing

\bibitem{austin2021programsynthesislargelanguage}
\BIBentryALTinterwordspacing
J.~Austin, A.~Odena, M.~Nye, M.~Bosma, H.~Michalewski, D.~Dohan, E.~Jiang, C.~Cai, M.~Terry, Q.~Le, and C.~Sutton, ``Program synthesis with large language models,'' 2021. [Online]. Available: \url{https://arxiv.org/abs/2108.07732}
\BIBentrySTDinterwordspacing

\bibitem{PanGenerativeAIForOCL}
F.~Pan, V.~Zolfaghari, L.~Wen, N.~Petrovic, J.~Lin, and A.~Knoll, ``Generative ai for ocl constraint generation: Dataset collection and llm fine-tuning,'' in \emph{2024 IEEE International Symposium on Systems Engineering (ISSE)}, 2024, pp. 1--8.

\bibitem{abukhalaf2023codexpromptengineeringocl}
\BIBentryALTinterwordspacing
S.~Abukhalaf, M.~Hamdaqa, and F.~Khomh, ``On codex prompt engineering for ocl generation: An empirical study,'' 2023. [Online]. Available: \url{https://arxiv.org/abs/2303.16244}
\BIBentrySTDinterwordspacing

\bibitem{diroccoUseLargeLanguage2025}
J.~Di~Rocco, D.~Di~Ruscio, C.~Di~Sipio, P.~T. Nguyen, and R.~Rubei, ``On the use of large language models in model-driven engineering,'' \emph{Software and Systems Modeling}, Jan. 2025.

\bibitem{dongSurveyIncontextLearning2024}
Q.~Dong, L.~Li, D.~Dai, C.~Zheng, J.~Ma, R.~Li, H.~Xia, J.~Xu, Z.~Wu, B.~Chang, X.~Sun, L.~Li, and Z.~Sui, ``A {{Survey}} on {{In-context Learning}},'' in \emph{Proceedings of the 2024 {{Conference}} on {{Empirical Methods}} in {{Natural Language Processing}}}.\hskip 1em plus 0.5em minus 0.4em\relax Miami, Florida, USA: Association for Computational Linguistics, 2024, pp. 1107--1128.

\bibitem{weiChainThoughtPromptingElicits2022}
J.~Wei, X.~Wang, D.~Schuurmans, M.~Bosma, B.~Ichter, F.~Xia, E.~Chi, Q.~Le, and D.~Zhou, ``Chain-of-{{Thought Prompting Elicits Reasoning}} in {{Large Language Models}},'' 2022.

\bibitem{yaoReActSynergizingReasoning2022}
S.~Yao, J.~Zhao, D.~Yu, N.~Du, I.~Shafran, K.~Narasimhan, and Y.~Cao, ``{{ReAct}}: {{Synergizing Reasoning}} and {{Acting}} in {{Language Models}},'' 2022.

\bibitem{UN157}
\BIBentryALTinterwordspacing
``Un regulation no.157 – uniform provisions concerning the approval of vehicles with regards to automated lane keeping systems [2021/389],'' pp. 75--137, Mar 2021. [Online]. Available: \url{http://data.europa.eu/eli/reg/2021/389/oj}
\BIBentrySTDinterwordspacing

\bibitem{fremontScenicLanguageScenario2023}
D.~J. Fremont, ``Scenic: A language for scenario specification and data generation,'' \emph{Machine Learning}, 2023.

\bibitem{joelSurveyLLMbasedCode2024}
S.~Joel, J.~J. Wu, and F.~H. Fard, ``A {{Survey}} on {{LLM-based Code Generation}} for {{Low-Resource}} and {{Domain-Specific Programming Languages}},'' 2024.

\bibitem{zolfaghariAdoptingRAGLLMAided2024}
V.~Zolfaghari, N.~Petrovic, F.~Pan, K.~Lebioda, and A.~Knoll, ``Adopting {{RAG}} for {{LLM-Aided Future Vehicle Design}},'' in \emph{2024 2nd {{International Conference}} on {{Foundation}} and {{Large Language Models}} ({{FLLM}})}.\hskip 1em plus 0.5em minus 0.4em\relax Dubai, United Arab Emirates: IEEE, Nov. 2024, pp. 437--442.

\bibitem{pan2025automatingautomotivesoftwaredevelopment}
\BIBentryALTinterwordspacing
F.~Pan, Y.~Song, L.~Wen, N.~Petrovic, K.~Lebioda, and A.~Knoll, ``Automating automotive software development: A synergy of generative ai and formal methods,'' 2025. [Online]. Available: \url{https://arxiv.org/abs/2505.02500}
\BIBentrySTDinterwordspacing

\end{thebibliography}

\clearpage
\appendices

\section{Example prompt for the Vehicle Definition pipeline}
\label{appendix:vehicle_def_prompt}

Below is an example prompt with two Chain-of-Thought (CoT) examples.
The prompt is parametrized with several items, which are omitted for clarity:
\begin{itemize}
    \item \emph{vehicle\_definition} - requirements that describe the vehicle's sensors
    \item \emph{schema} - JSON schema that defines the configuration structure
    \item \emph{blueprints} - available CARLA blueprints
    \item \emph{example\_requirements} - an example set of requirements
    \item \emph{example\_config} - an example JSON configuration that corresponds to the example requirements
\end{itemize}

\begin{lstlisting}[breaklines=true, frame=single, basicstyle=\tiny]
You are provided with a list of requirements that describe the ego vehicle, enclosed in the "vehicle_definition" XML tag:

<vehicle_definition>
{vehicle_definition}
</vehicle_definition>

You are also provided with a JSON schema that describes the structure of the JSON configuration files, enclosed in the "schema" XML tag:

<schema>
{schema}
</schema>

You are also provided a list of available vehicle blueprints:

<blueprints>
{blueprints}
</blueprints>

Here's an example set of requirements that describe the vehicle and the corresponding configuration file:

<example_requirements>
{example_requirements}
</example_requirements>

<example_config>
{example_config}
</example_config>

Here are also several examples that may help you translate certain requirements into configuration:

<example1>
Requirements:
[2] The ego vehicle is 5.5m long (X axis), 2.2m wide (Y axis), and 2.0m high (Z axis).
[4] The front cameras should be mounted centrally behind the windshield, at 1.3m height and 1m from the front.
Steps to solve the problem:
Sensor position is relative to the vehicle position, which is understood as a point in the middle of the bottom face of the vehicle's bounding box.
The bounding box in our case has the following dimensions: x = 5.5m, y = 2.2m, z = 2.0m.
This means that the sensors can be placed in the following ranges: x - between -2.75m (back) and 2.75m (front);
y - between -1.1m (left) and 1.1m (right); z - between 0m (bottom) and 2.0m (top).
To calculate the sensor positions:
- x position - The requirement tells us that the camera should be placed 1m from the front, which gives us 2.75m - 1m = 1.75m.
- y position - the camera is mounted centrally, so the y value should be 0.
- z position - the z=0 position of the sensor is at the bottom of the bounding box (the ground), so we must add the whole 1.3m.
In the end, we have:
"transform": {{ "x": 1.75, "y": 0.0, "z": 1.3}}
</example1>

<example2>
Requirements:
The rear camera should have a framerate of 10 FPS.
Steps to solve the problem:
The available camera property that corresponds to this requirement is "sensor_tick".
The property defines the time between two consecutive frames, which is the inverse of the desired FPS value.
In our case, the formula will be the inverse of 10FPS =  1s / 10 frames = 0.1 seconds per frame
In the end, we have:
"sensor_tick": 0.1
</example2>

Given the above, create a new JSON configuration file containing all relevant information in the requirements.
Return only the new configuration code, no additional text.
\end{lstlisting}

\section{Example requirements and configuration code for the Vehicle Definition pipeline}
\label{appendix:vehicle_def_basic}

The listing below presents the dataset used in experiment \ref{sec:vehicle_def_1}, visualized in Fig. \ref{fig:vehicle_def_sensors}.

\begin{lstlisting}[breaklines=true, frame=single, basicstyle=\tiny]
[1] The vehicle should be identified as "ego".
[2] The ego vehicle should be a Tesla Model 3.
[3] The ego vehicle is 4.7m long (X axis), 1.8m wide (Y axis), and 1.4m high (Z axis).

[4] The ego vehicle should have the following sensors mounted: a lidar, a mid-range camera, and a short-range camera.

[5] The lidar should be facing front.
[6] The lidar should have a range of 20 m.
[7] The lidar should have a horizontal field of view of 110 degrees.
[8] The lidar should have a vertical field of view of 20 degrees.
[9] The lidar should be placed centrally at the front of the vehicle, at a height of 0.5m.

[10] The mid-range camera should be facing front.
[11] The mid-range camera should have a range of 20 m.
[12] The mid-range camera should have a horizontal field of view of 50 degrees.
[13] The mid-range camera should have a vertical field of view of 35 degrees.
[14] The mid-range camera should have a resolution of at least 2 Mpx.
[15] The mid-range camera should be placed centrally, 1m from the front, at a height of 1.2m.

[16] The short-range camera should be facing front.
[17] The short-range camera should have a range of 10 m.
[18] The short-range camera should have a horizontal field of view of 120 degrees.
[19] The short-range camera should have a vertical field of view of 90 degrees.
[20] The short-range camera should have a resolution of at least 3 Mpx.
[21] The short-range camera should be placed centrally, 1m from the front, at a height of 1.2m.

[22] The rear camera should be facing the rear.
[23] The rear camera should have a range of 10 m.
[24] The rear camera should have a horizontal field of view of 120 degrees.
[25] The rear camera should have a vertical field of view of 90 degrees.
[26] The rear camera should have a resolution of at least 3 Mpx.
[27] The rear camera should be placed centrally at the back of the vehicle, at a height of 1.2m.
\end{lstlisting}

The listing below shows the LLM-generated configuration code corresponding to the above base requirements.

\lstset{
    string=[s]{"}{"},
    stringstyle=\color{blue},
    comment=[l]{:},
    commentstyle=\color{black},
}

\begin{lstlisting}[language=python, breaklines=true, frame=single, basicstyle=\tiny]
{
    "id": "ego",
    "blueprint": "vehicle.tesla.model3",
    "sensors": [
        {
            "id": "lidar_front",
            "blueprint": "sensor.lidar.ray_cast",
            "transform": {
                "x": 2.35,
                "y": 0.0,
                "z": 0.5,
                "pitch": 0.0,
                "yaw": 0.0,
                "roll": 0.0
            },
            "attributes": {
                "range": 20,
                "horizontal_fov": 110,
                "upper_fov": 10,
                "lower_fov": -10
            }
        },
        {
            "id": "mid_range_camera",
            "blueprint": "sensor.camera.rgb",
            "transform": {
                "x": 1.35,
                "y": 0.0,
                "z": 1.2,
                "pitch": 0.0,
                "yaw": 0.0,
                "roll": 0.0
            },
            "attributes": {
                "range": 20,
                "image_size_x": 2000,
                "image_size_y": 1000,
                "horizontal_fov": 50,
                "vertical_fov": 35,
                "sensor_tick": 0.05
            }
        },
        {
            "id": "short_range_camera",
            "blueprint": "sensor.camera.rgb",
            "transform": {
                "x": 1.35,
                "y": 0.0,
                "z": 1.2,
                "pitch": 0.0,
                "yaw": 0.0,
                "roll": 0.0
            },
            "attributes": {
                "range": 10,
                "image_size_x": 3000,
                "image_size_y": 2000,
                "horizontal_fov": 120,
                "vertical_fov": 90,
                "sensor_tick": 0.1
            }
        },
        {
            "id": "rear_camera",
            "blueprint": "sensor.camera.rgb",
            "transform": {
                "x": -2.35,
                "y": 0.0,
                "z": 1.2,
                "pitch": 0.0,
                "yaw": 180.0,
                "roll": 0.0
            },
            "attributes": {
                "range": 10,
                "image_size_x": 3000,
                "image_size_y": 2000,
                "horizontal_fov": 120,
                "vertical_fov": 90,
                "sensor_tick": 0.1
            }
        }
    ]
}
\end{lstlisting}

The listing below presents an extension for the dataset used in experiment \ref{sec:vehicle_def_1}, visualized in Fig. \ref{fig:vehicle_def_sensors_extended}.

\begin{lstlisting}[breaklines=true, frame=single, basicstyle=\tiny]
[28] The ego vehicle should have a left rearward-looking side camera.
[29] The left rearward-looking side camera should have a horizontal field of view of 70 degrees.
[30] The left rearward-looking side camera should face 120 degrees to the vehicle's left.
[31] The left rearward-looking side camera should be placed on the left side of the vehicle, 1m from the front, at a height of 1m.
[32] The left rearward-looking side camera should have a resolution of 1448 x 938.

[33] The ego vehicle should have a right rearward-looking side camera.
[34] The right rearward-looking side camera should have a horizontal field of view of 70 degrees.
[35] The right rearward-looking side camera should be facing 120 degrees to the vehicle's right.
[36] The right rearward-looking side camera should be placed on the right side of the vehicle, 1m from the front, at a height of 1m.
[37] The right rearward-looking side camera should have a resolution of 1448 x 938.
\end{lstlisting}

\section{Example requirements for the Vehicle Definition pipeline}
\label{appendix:vehicle_def_req}

The listing below contains the requirement dataset used in sections \ref{sec:vehicle_def_2} and \ref{sec:vehicle_def_3}.
\begin{lstlisting}[breaklines=true, frame=single, basicstyle=\tiny]
[1] The ego vehicle should be a Tesla Model 3.
[2] The ego vehicle is 4.7m long (X axis), 1.8m wide (Y axis), and 1.4m high (Z axis).

[3] The ego vehicle should have the following forward cameras mounted: the wide camera, the main camera, and the narrow camera.
[4] The front cameras should be mounted centrally behind the windshield, at 1.3m height and 1m from the front.
[5] The front cameras should be facing front.
[6] The front cameras should have a resolution of 2896 x 1876.

[7] The wide front camera should have a horizontal field of view of 120 degrees.
[8] The main front camera should have a horizontal field of view of 70 degrees.
[9] The narrow front camera should have a horizontal field of view of 50 degrees.

[10] The ego vehicle should have left and right forward-looking side cameras.

[11] The left forward-looking side camera should have a horizontal field of view of 90 degrees.
[12] The left forward-looking side camera should be facing 60 degrees to the vehicle's left.
[13] The left forward-looking side camera should be placed in the middle of the left side of the vehicle, at a height of 1m.
[14] The left forward-looking side camera should have a resolution of 1448 x 938.

[15] The right forward-looking side camera should have a horizontal field of view of 90 degrees.
[16] The right forward-looking side camera should be facing 60 degrees to the vehicle's right.
[17] The right forward-looking side camera should be placed in the middle of the right side of the vehicle, at a height of 1m.
[18] The right forward-looking side camera should have a resolution of 1448 x 938.

[19] The ego vehicle should have the following rearward-looking side cameras: left and right.

[20] The left rearward-looking side camera should have a horizontal field of view of 70 degrees.
[21] The left rearward-looking side camera should be facing 120 degrees to the vehicle's left.
[22] The left rearward-looking side camera should be placed on the left side of the vehicle, 1m from the front, at a height of 1m.
[23] The left rearward-looking side camera should have a resolution of 1448 x 938.

[24] The right rearward-looking side camera should have a horizontal field of view of 70 degrees.
[25] The right rearward-looking side camera should be facing 120 degrees to the vehicle's right.
[26] The right rearward-looking side camera should be placed on the right side of the vehicle, 1m from the front, at a height of 1m.
[27] The right rearward-looking side camera should have a resolution of 1448 x 938.

[28] The ego vehicle should have a rear camera.
[29] The rear camera should be mounted centrally, at the back of the vehicle, at a height of 1m.
[30] The rear camera should have a horizontal field of view of 120 degrees.
[31] The rear camera should be facing back.
[32] The rear camera should have a resolution of 1448 x 938.

[33] All cameras should have a framerate of 24 FPS.
\end{lstlisting}

\section{Example requirements and agent placement code for the Pre-conditions pipeline}
\label{appendix:precondition_req}

The listing below presents an example car-to-car scenario extracted from the UN Standard No. 152 \cite{UN152}, used in section \ref{sec:pre_conditions1}.

\begin{lstlisting}[breaklines=true, frame=single, basicstyle=\tiny]
[1] Two vehicles of Category M1 AA saloon shall be present: the subject vehicle that performs the braking and the lead vehicle
[2] Both vehicles should face in the same direction of travel
[3] The subject vehicle shall approach the lead vehicle in a straight line for at least 2 seconds before the functional part of the test commences
[4] The subject vehicle should travel at a speed of 20 km/h when the vehicle brakes
[5] The functional part of the test shall start at a distance corresponding to a Time To Collision (TTC) of at least 4 seconds from the target
[6] The subject vehicle is unladen
[7] The lead vehicle is stationary
[8] The road is dry
[9] Ambient illumination is at least 1000 lux
\end{lstlisting}

The listing below presents an example of Python code used for agent placement.
The code was generated based on the requirements in the listing above.

\begin{lstlisting}[language=Python, breaklines=true, frame=single, basicstyle=\tiny]
def position_agents(graph: nx.Graph) -> list:
    # Define speed for the subject vehicle in km/h (requirement 4)
    subject_speed_km = 20
    subject_speed = subject_speed_km * 1000 / 3600  # Convert to meters per second

    # The lead vehicle is stationary (requirement 7)
    lead_speed = 0

    # Calculate the distance that the subject vehicle is going to cover before the functional part of the test (requirement 3)
    time_before_test = 2  # In seconds
    subject_dist_before_test = subject_speed * time_before_test

    # Calculate the distance for the functional part of the test (requirement 5)
    ttc = 4  # Time-to-Collision in seconds
    dist_during_test = ttc_to_distance(ttc, subject_speed, lead_speed)

    # Calculate the minimal length of the route
    route_min_length = subject_dist_before_test + dist_during_test

    # Retrieve all relevant routes (requirement 3)
    routes = get_routes_straight(graph)

    # Find all routes that match the distance requirements
    routes_filtered = filter_routes_by_length(routes, route_min_length)
    if len(routes_filtered) == 0:
        raise Exception("No route matches the distance requirements")

    selected_route = routes_filtered[0]

    # Position the subject vehicle at the beginning of the route, facing forward (requirement 2)
    subject_transform = create_spawnpoint(selected_route, 0, AgentRotation.FORWARD)

    # Position the lead vehicle at the specified distance for the functional part, using backward for no effect since it's stationary (requirement 7)
    lead_transform = create_spawnpoint(selected_route, subject_dist_before_test + dist_during_test, AgentRotation.FORWARD)

    # Select the target locations for both vehicles
    # The subject vehicle should be driving towards the end of the selected route
    subject_target = selected_route[-1]

    # The lead vehicle is stationary, its target position is the same as its initial position
    lead_target = lead_transform.location

    return [selected_route, subject_transform, subject_target, lead_transform, lead_target]
\end{lstlisting}

The listing below presents an example car-to-pedestrian scenario extracted from the UN Standard No. 152 \cite{UN152}, used in section \ref{sec:pre_conditions1}.

\begin{lstlisting}[breaklines=true, frame=single, basicstyle=\tiny]
[1] A pedestrian target is crossing the road at a constant speed of 5km/h, perpendicular to the trajectory of the subject vehicle.
[2] The pedestrian target positioning shall be coordinated with the subject vehicle so that the impact point of the pedestrian target on the front of the subject vehicle is on the longitudinal centerline of the subject vehicle.
[3] The subject vehicle shall approach the point of collision in a straight line for at least 2 seconds before the functional part of the test commences.
[4] The subject vehicle should travel at 20 km/h when the vehicle brakes.
[5] The functional part of the test shall start at a distance corresponding to a Time To Collision (TTC) of at least 4 seconds from the target.
[6] The subject vehicle is unladen.
[8] The road is dry.
[9] Ambient illumination is at least 1000 lux.
\end{lstlisting}

\section{Example requirements and configuration code for the Post-conditions pipeline}
\label{appendix:postcondition_req}

The listing below presents an example car-to-pedestrian scenario extracted from the UN Standard No. 152 \cite{UN152}, used in section \ref{sec:post_conditions1}.

\begin{lstlisting}[breaklines=true, frame=single, basicstyle=\tiny]
[1] The subject vehicle should travel at a speed of 20 km/h when the vehicle brakes
[2] The braking demand was at least 5m/s^2
[3] The ego vehicle does not crash into the lead vehicle
[4] The end speed of the ego vehicle is 0km/h
\end{lstlisting}

The listing below shows the LLM-generated configuration code corresponding to the base requirements.

\begin{lstlisting}[language=python, breaklines=true, frame=single, basicstyle=\tiny]
{
    "telemetry": [
        {
            "id": "ID_TARGET_SPEED",
            "sensor": "speed",
            "begin": "reached_target_speed",
            "end": null,
            "operator": "==",
            "value": 20
        },
        {
            "id": "ID_BRAKING_FORCE",
            "sensor": "brake",
            "begin": "braking_start_aeb",
            "end": "braking_end_aeb",
            "operator": ">=",
            "value": 5
        },
        {
            "id": "ID_COLLISION",
            "sensor": "collision",
            "begin": "simulation_start",
            "end": "braking_end_aeb",
            "operator": "==",
            "value": false
        },
        {
            "id": "ID_END_SPEED",
            "sensor": "speed",
            "begin": "braking_end_aeb",
            "end": null,
            "operator": "==",
            "value": 0
        }
    ]
}
\end{lstlisting}

The \emph{sensor} field specifies the measured quantity type.
These quantities are predefined and linked to CARLA’s internal telemetry data.

The \emph{begin} and \emph{end} fields indicate the time window during which the measurement should take place.
Since it's often challenging—or even unfeasible—to define this interval using absolute time, predefined events are used instead.
For example, the \emph{braking\_star\_aeb} event is triggered when the AEB system issues a braking command.
If the \emph{end} field is left empty (null), the check is carried out at a single point in time defined by the \emph{begin} event.

The \emph{operator} and \emph{value} field determine how the quantity is evaluated.
For instance, in the \emph{ID\_BRAKING\_FORCE} check, the automatic braking demand must be at least $5m/s^2$.
This is represented by setting the \emph{operator} to $>=$ and the \emph{value} to $5$.

\end{document}